\documentstyle[preprint,version2,aps]{revtex}

\begin{document}
\draft
\begin{title}
\begin{bf}
Critical Properties of Quantum Many-Body Systems \\
with $1/r^2$ Interaction
\end{bf}
\end{title}
\author{Norio Kawakami }
\begin{instit}
Yukawa Institute for Theoretical Physics, \\
Kyoto University, Kyoto 606
\end{instit}
\begin{abstract}
We review recent results obtained for a class of
one-dimensional quantum models
with $1/r^2$ long-range interaction.
Based on the asymptotic Bethe-ansatz solution and
conformal field theory, we study critical properties of the
continuum boson model,
the SU($\nu$) spin chain, the
OSp($\nu$,1) supersymmetric {\it t-J} model,
and a new hierarchy of models
related to the fractional quantum Hall effect.
We further investigate  the class of $1/r^2$
models with  harmonic confinement
by means of a newly proposed method
of the {\it renormalized-harmonic oscillator}
solution.
\end{abstract}
\vspace{1 cm}    
\pacs{PACS numbers: }
\narrowtext


\noindent
\begin{center}
\begin{bf}
 \S 1. Introduction
\end{bf}
\end{center}

\vskip 2mm

One-dimensional (1D) many-body systems with
long-range interaction of $1/r^2$ type \cite{calo,sutha,suthb}
have been studied extensively in connection
with fundamental notions in condensed
matter and statistical physics.
Even in classical models \cite{moser,ferro,bak,pere}, one
can find remarkable results.  For example,
a path-integral formulation of
the Kondo problem results in
the ferromagnetic Ising model with $1/r^2$ interaction \cite{ferro}.
Also, the antiferromagnetic Ising model shows
a typical example of the devil staircase \cite{bak}.

Quantum models with $1/r^2$ interaction
have attracted renewed interest considerably [1-3, 8-42],
revealing new interesting aspects of integrable systems,
and thus providing us with a paradigm
of fundamental ideas actively discussed
in condensed matter physics,
such as the random matrix \cite{sutha,suthb},
the Gutzwiller state for
correlated electrons \cite{hala,shastry,kuramoto},
the fractional quantum Hall
effect (FQHE) \cite{halc,kawac,poly,for,kawad,iso},
the level statistics for disordered
systems \cite{alt,narayan,zirn}, etc.

A common feature in quantum
$1/r^2$ systems is that the ground state is exactly given by
a Jastrow-type wavefunction [1-3, 8-22], namely a product of two-body
functions.  This characteristic nature of the wavefunction
should be  closely related to the integrability of
$1/r^2$ models because it implies that
the two-body scattering is essential in spite of the
long-range interaction.
More recently, the integrability of the class of quantum
Hamiltonians has been shown, and
algebraic structures for the underlying symmetry have been
clarified systematically [26-37].

The asymptotic Bethe-ansatz solution (ABA)
provides us with a systematic way to construct the energy
spectrum \cite{sutha,suthb}, which has been known to
give the exact solution to the continuum boson model and
also to the SU(2) spin chain \cite{halc}.
More recently, the ABA solution has been systematically
generalized to multicomponent quantum systems with
$1/r^2$ interaction, such as the SU($\nu$) spin chain,
the OSp($\nu$,1) supersymmetric {\it t-J} model,
and a new hierarchy of the models
related to the FQHE \cite{kawaa,kawab,kawac,kawad}.
Furthermore, it has been found that a family of
confined models with  harmonic potential
can be solved  by the {\it renormalized harmonic oscillator}
(RHO) method \cite{kawag}, which is a variant of the ABA.

In this paper  we wish to give a brief review of
our recent studies on the critical properties
of quantum $1/r^2$ models based on the
ABA solution and conformal field theory (CFT).
This paper is organized as follows. In the next section we
outline the ABA method by taking Sutherland's boson model
as a simple example \cite{sutha,suthb},
and then discuss its critical properties
based on CFT \cite{kya}. In \S 3 the ABA solution is further
extended to multicomponent systems with more complicated
internal symmetry, the SU($\nu$) Haldane-Shastry
spin chain \cite{kawab},
 and then in \S 4 the effects of hole-doping  are discussed by
the OSp($\nu,1$) supersymmetric {\it t-J} model \cite{kawaa,kawab}
which was introduced first by Kuramoto and Yokoyama for the
OSp(2,1) case \cite{kuramoto}.
We further introduce in \S 5 a new family of $1/r^2$ models
which are closely related to a certain hierarchy of
the FQHE \cite{kawac,kawad}.
In \S 6, we propose a new approach  based on
the RHO method  \cite{kawag} in order to
systematically construct the energy spectrum
for the models with harmonic confinement
\cite{kawag,kawah,vacek,vaceka,kawakura}.
We then prove, by explicitly
constructing the eigenfunctions for the
grond state \cite{vacek} as well as the excited states
 \cite{vaceka}, that the RHO indeed provides the
exact spectrum for the above systems.
Some applications to mesoscopic systems are also
mentioned.


\vskip 10mm
\begin{center}
\begin{bf}
  \S 2. Critical Properties of Continuum Boson Model
\end{bf}
\end{center}
\vskip 3mm

There are several variants of the integrable quantum models with
$1/r^2$  interaction.  In this paper we will
systematically investigate the systems with
periodic boundary conditions and also
the confined systems with harmonic potential. We note that
quantum $1/r^2$  models in continuum space
was introduced by Calogero many years ago \cite{calo},
and then have been actively studied by Sutherland \cite{sutha,suthb}.
In this section, we wish to  mention
characteristic properties of the quantum $1/r^2$ systems by
taking the boson model with periodic boundary conditions, and
outline how to apply the ABA method.

Let us introduce $N$ interacting bosons in a 1D chain
of circumference $L$,
\begin{equation}
{\cal H}= -\sum_{j=1}^N {\partial^2 \over \partial r^2_j}
+ \sum_{i<j} U(r_i-r_j),
\end{equation}
with  $1/r^2$ interaction $U(r_i-r_j)$.
In order to consider the model with  periodic boundary conditions,
the following form of the interaction should be used \cite{sutha},
\begin{equation}
U(r) =V\sum_{n=-\infty}^\infty (r+nL)^{-2}=
{V\pi^2 \over L^2}  \sin ^{-2}({\pi r \over L})
\equiv  V d_{ij}^{-2},
\end{equation}
which leads to $V/r^2$ in the limit of $L \rightarrow \infty $.
Here $d_{ij}$ corresponds to the chord distance for the ring.
It is known that the ground state for the model
is given by the Jastrow wavefunction,
i.e. the product of two-body functions \cite{calo,sutha},
\begin{equation}
\psi= \prod_{j<l} \, | \sin {\pi(r_j-r_l)\over L}|^\lambda
\end{equation}
with the Jastrow parameter,
\begin{equation}\lambda=[\sqrt{1+2V} +1]/2.
\end{equation}
We will be concerned with the repulsive case $V \geq 0$ hereafter.
It should be noted that all the integrable $1/r^2$ models
take the Jastrow wavefunction as the ground state
wavefunction.  The two-body nature of the ground state
is related to the integrability of the system, and
implies that the two-body scattering may be essential
for the many-body scattering. Based on this observation,
Sutherland proposed the ABA method to
exactly construct the excitation spectrum and
full thermodynamics  of the above boson system
\cite{sutha}.

\vskip 7mm
\noindent
{\it 2.1.  Asymptotic Bethe-ansatz }
\vskip 2mm

An asymptotic Bethe-ansatz (ABA) solution
provides an elegant method to construct the
excitation spectrum for the class of the above Hamiltonians
\cite{sutha}. The essence of the idea is that although
ordinary Bethe-ansatz (BA) methods are not applicable
 to the above systems with
long-range interactions,  the eigenfunctions
can be written down consistently like those for the BA,
\begin{equation}
\psi= \sum_{P} A(Q;P) \exp[i \sum_j k_{Pj} x_{Qj} ],
\end{equation}
in the {\it asymptotic} region,
$x_{Q1} \ll x_{Q2} \ll
 \cdots \ll x_{QN}$,  where $Q$ ($P$) expresses one
of $N!$ permutations for the coordinate (momentum)
configurations.  This form of the wavefunction implies  that
the  many-body $S$-matrix in the asymptotic region
can be {\it factorized} into two-body matrices
consistently. If this is  the case, one can diagonalize the
many body scattering based on the
factorized $S$-matrix, and then obtain the
spectrum of the system. The factorization of the
$S$-matrix may be somehow suspected from the two-body nature of the
Jastrow wavefunction for  the ground state.
At first glance the ABA solution seems to work only for
low-energy excitations of
continuum systems with the low density of particles.
Remarkably enough, however, it turns out that the ABA solution
{\it exactly} reproduces the whole  energy spectrum for
 boson systems of any density
\cite{sutha}.  Furthermore,  this method
has been found to be also applicable for the  lattice models with
high  density of particles \cite{halc}.

The two-body scattering in the Sutherland model
 yields the  $S$-matrix, $S_{ij}=-\exp[-i\theta(k_i-k_j)]$ with
the phase shift function
 $\theta(k)=\pi(\lambda-1) {\rm  sgn}(k)$.
Imposing periodic boundary conditions, one can
now diagonalize the many-body $S$-matrix
using the factorized $S$ matrices, and then deduce
the ABA equation for the rapidity $k_j$
\cite{sutha},
\begin{equation}
k_jL=2\pi I_j +  (\lambda-1)
\sum_{l=1}^N \Phi(k_j-k_l)
\end{equation}
with $\Phi(k)= \pi {\rm sgn}(k_j-k_l)$,
where $I_j$ is the quantum number
which satisfies the selection rule
$I_j=(N+1)/2 $ mod 1. The interaction effects
are now incorporated
into $k_j$ via the phase shift function,
and the energy is expressed simply in the
noninteracting form, $E=\sum k_j^2$.
Sutherland studied full thermodynamics based on the above
ABA equations \cite{sutha}. More recently, the critical behavior of
correlation functions have been clarified
\cite{kya} with the help of CFT
\cite{bpz,bcn,cardy,review}.

\vskip 7mm
\noindent
{\it 2.2. Conformal properties}
\vskip 2mm

By combining the ABA solution with
the finite-size scaling in CFT,
we now study  critical properties of the model,
and then evaluate the critical exponents
of correlation functions.
We briefly review the results obtained in  ref. \cite{kya}.
Let us start with low-temperature properties
of the free energy \cite{sutha},
\begin{equation}
F(T) \simeq F(T=0) -{\pi T^2 \over 6v} \ ,
\end{equation}
where  the velocity of elementary excitations is
$v=2\pi\lambda n$. According to the known
formula for finite-size scaling
in CFT \cite{bcn},  we can read the central charge of the underlying
Virasoro algebra as $c=1$ \cite{kya}.  Therefore, the critical
behavior of the present model is expected to be
described by $c=1$ Gaussian CFT.
To confirm this, we next calculate
the finite-size corrections to  excited states.
Elementary excitations can be specified by the deviation of
quantum numbers from the ground-state distribution.
If we take $I_j=(2j-N-m-1)/2+d$, the quantum number $m $
labels  the excitation which  changes the particle number,
whereas $d$ denotes the excitation
which carries the large momentum  $2k_Fd$ with $k_F=\pi n$.
We then classify the excitation as \cite{kya},
\begin{equation}
\Delta E \simeq {2\pi v \over L} x(m;d;n^\pm ),
\end{equation}
where the chemical potential term has been omitted.
According to the finite size scaling in CFT \cite{cardy},
we can deduce the scaling dimension,
\begin{equation}x(m;d;n^\pm )= {\lambda \over 4}
m^2+{1 \over \lambda}d^2 +n^+ +n^- \ ,
\end{equation}
where particle-hole excitations are denoted by
 non-negative integers  $n^\pm $.
The selection rule for the quantum number reads:
 $d$=integer for bosons.
The momentum carried by the above excitation is
\begin{equation}P=2\pi k_F d+ {2\pi \over L}
[m d+ n^+-n^-] \ .
\end{equation}
{}From (9) and (10) we can determine conformal weights $\Delta^\pm $
for the holomorphic (antiholomorphic) piece,
which characterize the operator content of the underlying
Virasoro algebra in CFT \cite{bpz}.
Recalling that CFT predicts  the energy
and the momentum to be universally related to conformal weights
as $x=\Delta^+ +\Delta^-$
and $P=(2\pi/L)(\Delta^+ -\Delta^-)$
for the $1/L$ sector \cite{cardy},
 we obtain conformal weights \cite{kya},
\begin{equation}
\Delta^\pm (m;d;n^\pm ) = {1 \over 2}
\big(  {m \over 2R}\pm  d R \big)^2 + n^\pm
\end{equation}
with $R=1/\sqrt{\lambda}$.
We now confirm that the above
expression for conformal weights is inherent in
$c=1$ Gaussian CFT realized by
free bosons with periodicity $R$, in which
non-negative integers $n^\pm $ feature the conformal
tower \cite{review}.
{}From the above analysis \cite{kya}, we now conclude
that the present model with $1/r^2$ interaction
is a typical example of Luttinger liquids \cite{halb,fk,kyc},
i.e. its critical behavior is controlled by
$c=1$ CFT.

\vskip 7mm
\noindent
{\it 2.3. Correlation exponents }
\vskip 2mm

It is now easy to determine the critical exponents of
correlation functions. For instance, let us begin with
long-distance behavior of
the  density correlation function,
 \begin{equation}
<n (r)n (0)> \simeq  {\rm const.} + a_0r^{-2}
       + a_2 r^{-\alpha} \cos 2k_F r  \ ,
 \end{equation}
where $n(r)$ is the density operator.
By taking $(m, d, n^\pm ) = (0, 1, 0) $
as the quantum numbers, which carry the $2k_F$
momentum transfer,  we obtain
the critical exponent ($\alpha=2x$),
\begin{equation}
\alpha=2/\lambda
\end{equation}
for $2k_F$  oscillation piece \cite{kya}.
Note that there is no logarithmic correction to
the correlation function. On the other hand,
the field correlator  for bosons
 \begin{equation}<\phi_b^{\dag}(r) \phi_b(0)> \simeq
r^{-\beta_b} \ ,
\end{equation}
has the leading non-oscillation term.
By taking the set of quantum numbers
 $(m, d, n^\pm )=(1, 0, 0)$ for the primary field,
we obtain the corresponding critical exponent \cite{kya},
\begin{equation}
\beta_b=\lambda/2.
\end{equation}
Fourier transform of this correlator
yields the momentum distribution function
around the origin,
$ n_b(k) \simeq |k|^{\theta _b}$,
with the corresponding critical exponent
$\theta _b=\beta_b-1=\lambda/2-1$.

As the interaction strength $V$  increases,
the critical exponents
vary  continuously, characterizing the U(1)
CFT critical line.
For the special values such as $V= 0$ and $ 4$,
the present exponents \cite{kya} agree with those obtained by
Sutherland using   the random matrix theory \cite{sutha}.
We note that
the system exhibits  interesting properties
at $V=4$, at which
the periodicity of bosons becomes $R=1/\sqrt{2}$.
This implies that symmetry of the model
is enhanced to SU(2),  and is
described by the level-1 SU(2) Wess-Zumino-Witten
model \cite{review}.
Therefore the effective theory for  this special point
is equivalent to that for
the SU(2) Haldane-Shastry spin chain \cite{hala,shastry}.

The above critical exponents satisfy
the universal scaling relations for the Luttinger liquid,
$\beta_b=1/\alpha$ \cite{halc}.  It is instructive to note that
in contrast to the known models with short-range interaction,
the above critical exponents do not depend on the
density of particles \cite{hala,kya}, but only on the
interaction strength $V$.  This peculiar property
is characteristic of $1/r^2$ quantum models. We will encounter
similar examples in the following chapters.

We have been concerned with the boson case so far.
The CFT analysis for the fermion case
can be performed similarly \cite{kya}.
For fermions, the selection rule should read
$d= m/2 \ \ {\rm mod} \ 1$, reflecting
antisymmetry nature of the wavefunction.
Hence the field correlator for fermions
has the leading $k_F$ oscillation term and
the corresponding exponent is given
as $\beta_f=[\lambda+1/\lambda]/2$ by choosing the
quantum numbers as $(m, d, n^\pm )=(1, 1/2, 0)$.
This leads to the critical exponent
of the  momentum distribution
around the Fermi point, $\theta _f=
[\lambda+1/\lambda-1]/2$.

\vskip 7mm
\noindent
{\it 2.4.  Haldane-Shastry spin chain}
\vskip 2mm

The quantum $1/r^2$ model was extended independently
by Haldane and Shastry to the lattice
case, namely the antiferromagnetic
$S=1/2$ spin chain with
$1/r^2$ exchange interaction \cite{hala,shastry}.
The Hamiltonian for a periodic ring with $L$ sites
reads,
\begin{equation} {\cal H} = \sum_{i<j} J_{ij}
[S_i^xS_j^x+S_i^yS_j^y + {1 \over 2}p(p-1) S_i^zS_j^z]   \ ,
\end{equation}
where  $J_{ij}$ is chosen
to satisfy periodic boundary conditions,
$J_{ij}= J d(x_i-x_j)^{-2}$
with the chord distance $ d(x) =(L/\pi)\sin (\pi x/ L)$
(the $i-$th site is denoted by $x_i$).
Here $p$ represents the anisotropy of the model
which is  assumed to be $p \geq 2$ in
what follows.  The isotropic model ($p=2$) has been
quite well investigated by Haldane and Shastry \cite{hala,shastry}.
Particularly in this case the ground-state
wavefunction is given by the completely-projected
Gutzwiller wavefunction at half filling \cite{hala,shastry},
\begin{equation}
\Psi_G(\{x_i\})
=  \exp(-i\pi \sum_{i}x_i)
  \prod_{i<j} d(x_i-x_j)^2,
\end{equation}
in terms of coordinates $\{x_i\}$ for down spins.
As for the anisotropic case,
the ground-state was also obtained
in the Jastrow form for a
positive even integer $p$ \cite{hala}. Here we briefly
summarize  CFT analysis of the anisotropic Haldane-Shastry
model \cite{kya}.

It is known that the ABA solution is
applicable to the above model
although  the asymptotic region is not realized for a lattice
system with high densities of particles \cite{halc}.
The two-body $S$-matrix for the model is
obtained as  $S_{ij}=-\exp[-i\theta(k_i-k_j)]$ with
the phase shift function $\theta(k)=(p-1) \pi {\rm  sgn}(k)$.
 Hence the ABA equation
is essentially the same as the continuum boson case \cite{halc},
\begin{equation}
k_jL=2\pi I_j + (p-1) \sum_{l=1}^N \Phi(k_j-k_l),
\end{equation}
where $N$ is the number of down spins.
In contrast to the continuum model,
however, the available range of $k_j$
is restricted to  $[-\pi,\pi]$, reflecting the
periodicity of the lattice.
One of the crucial consequences due to this restriction is
that there exists
the lower bound for the magnetization in the liquid phase,
$  s_z=1/2 -1/p$.
Therefore the massless phase is realized
for magnetic fields, $ H_{c1} \leq H \leq H_{c2}$, where
the upper critical field is given by
\begin{equation}
H_{c2}={\pi^2 J \over 12}[p(p-1)+1],
\end{equation}
at which the system is  fully polarized,
and  the lower critical field is
\begin{equation}
H_{c1}={\pi^2 J \over 6}[{1 \over 2}p(p-1)-1],
\end{equation}
at which the magnetization takes its minimum value
$s_z=1/2 -1/p$ in the liquid phase \cite{kya}.
The ABA solution is applicable only
for this range of magnetic fields.  We would like to
mention that the supermultiplet structure discovered
by Haldane \cite{halc},
which is now classified by the Yangian algebra \cite{halz},
made it possible to construct the full thermodynamics
for the {\it isotropic} case ($p=2$).

Similarly to the Sutherland model,
the low temperature free energy
in the liquid phase takes the form, $F(T) \simeq F(T=0) -\pi T^2/(6v)$
with the spin velocity $v=\pi p(1-2s_z)/(4J)$.
This implies  that the spin-liquid behavior of the present model
is described by CFT with
$c=1$. Also, the excitation spectrum
turns out to be classified in the same form as in (9).
Hence, we can conclude that the spin-liquid phase
for $ H_{c1} \leq H \leq H_{c2}$
is classified as that of  Luttinger liquids \cite{kya}.

Following the method outlined for boson case,
it is straightforward to deduce correlation exponents.
For instance, the asymptotic form of the spin correlation function
is written  as
\begin{equation}<S^z(x)S^z(0)> \simeq  c_0x^{-2}
       + c_2 x^{-\alpha} \cos 2k_F x  \ .
 \end{equation}
with the exponent $\alpha=2/p$.
On the other hand,  the transverse spin correlation
has the leading non-oscillation term
 \begin{equation}<S^+(x) S^-(0)> \simeq  x^{-\beta} \ .
\end{equation}
with  the exponent $\beta=p/2$.
These exponents indeed satisfy the universal scaling
relation for the Luttinger liquid, $\alpha=1/\beta$,
as is the case for the continuum case. Note that
conformal properties for
the isotropic case ($p=2$)
has already been discussed by Haldane in detail \cite{halc}.

We have been concerned  so far with the liquid phase
with $s_z\geq 1/2 -1/p$ under applied magnetic fields
$H_{c1}\leq H\leq H_{c2}$.  We then ask
what will happen for  $H<H_{c1}$.
Unfortunately the ABA solution is not efficient to answer
this question. We can say  that the system
shows a singularity at $H=H_{c1}$,
but it is not clear whether the ordered phase
is realized below $H_{c1}$ or not.  We close
this section by mentioning  key points to solve the
above question.  We first  point out that
when the system approaches the boundary $H_{c1}$
from the liquid phase, the spin correlation with the period $p$ is
enhanced at $H=H_{c1}$ \cite{kya}. So, it may be interesting to check
what kind of phase would be stabilized just below $H_{c1}$.
It is also instructive to ask
what happens for the magnetization when $H$ decreases further
below $H_{c1}$.  According to the results in
the Ising limit \cite{bak},
it may be possible that magnetization would
show stair structures as a function of $H$.
These interesting issues are
to be clarified in the future study.

\vskip 15mm
\begin{center}
\begin{bf}
\S 3.   SU($\nu$) Spin Chain
\end{bf}
\end{center}
\vskip 3mm

Now we wish to extend the quantum $1/r^2$ models to
multicomponent cases.
Such a generalization was first
made by Kuramoto and Yokoyama using the
supersymmetric {\it t-J} model, who found
 the Gutzwiller state as the
ground state  and discussed
low-energy excitations \cite{kuramoto}.  Subsequently,
this model was solved by the ABA method \cite{kawaa},
which has been later proven to give the exact
spectrum of the model \cite{wang}.
 Further generalization to the SU($\nu$)
spin chain and to the multicomponent {\it t-J}
model with OSp($\nu,1$) supersymmetry
has been done independently in refs. \cite{kawab,ha}.
Before discussing the supersymmetric {\it t-J} model
we first study  a SU($\nu$) generalization of the
Haldane-Shastry model ($\nu \geq 2$), and then
discuss the effects of hole-doping using the
supersymmetric {\it t-J} model in the next section.

We introduce the Hamiltonian of the
SU($\nu$) spin chain with $1/r^2$
interaction by the following general form \cite{kawab},
\begin{equation}
{\cal H}= {1 \over 2}
\sum_{i<j, \alpha, \beta}
(-1)^{F(\beta)} J_{ij}
X_i^{\alpha\beta}X_j^{\beta\alpha} ,
\end{equation}
where the Hubbard operator $X_i^{\alpha\beta}=|i\alpha><i\beta|$
interchanges  states at $i$-th site from $\beta$
to $\alpha$, and the
exchange coupling $J_{ij}$ is given by inverse-square
interaction (16).  Here we have assumed that the system consists
of $\nu$ components of spins (or colors) with $\alpha,
\beta=1, 2, \cdots, \nu$, and the  fermion numbers
take $F(\beta)=(0,0, \cdots, 0)$
for $\beta=(1,2,\cdots, \nu)$ in case of the
SU($\nu$) spin chain. Hence all the particles
obey the same statistics. This Hamiltonian,
which is a SU($\nu$) generalization of the Haldane-Shastry
model \cite{kawab,ha},
is indeed invariant under the global SU($\nu$) transformation.
If we add another state ($\alpha=\nu+1$) with
the different fermion number $F(\nu+1)=1$, this Hamiltonian describes
the multicomponent  {\it t-J} model with OSp($\nu,1$)
supersymmetry which is realized by
doping holes into the  SU($\nu$) Haldane-Shastry model.

\vskip 7mm
\noindent
{\it 3.1. Ground-state wavefunction}
\vskip 2mm

We start by writing down the completely projected
SU($\nu$) Gutzwiller wavefunction as the
exact ground state for the above SU($\nu$)
spin chain \cite{kawab,ha},
\begin{equation}
\ |\Psi_G>= P_G^{(1)} \prod_{\alpha=1}^{\nu}
\prod_{k_\alpha}^{k_F} a_{k_\alpha}^{(\alpha) \dag}|0>,
\end{equation}
where $a_{k_\alpha}^{(\alpha) \dag}$ is the creation
operator of electrons with spin $\alpha$
($ 1 \leq \alpha \leq \nu$) and momentum $k_\alpha$.
We have assumed here that  $P_G^{(1)}$
selects configurations for which every cite is occupied
by only a single electron with spin $\alpha$.
The  projection $P_G^{(1)}$ is easily
done by taking the reference state $|F>$  full of
particles with  $\nu$-th spin \cite{anderson}.
In this representation, we rewrite the
Gutzwiller state as,
\begin{equation}
 | \Psi_G>= \sum_{(\alpha,i)}
\Psi_G(\{x_i^{(\alpha)}\})
\prod_{\alpha, i} b_{i}^{(\alpha) \dag}|F>,
\end{equation}
where $b_{i}^{(\alpha) \dag}
=a_{i}^{(\alpha) \dag} a_{i}^{(\nu)}$
 a creation operator for spin particles
($  1 \le \alpha \le \nu-1$).  The Gutzwiller
wavefunction is given in terms of coordinates
of spin particles ($x_i^{(\alpha)}$),
\begin{equation}
\Psi_G(\{x_i^{(\alpha)}\})
=  \exp(-i\pi \sum_{\alpha,i}x_i^{(\alpha)})
  \prod_{\alpha, i<j} d(x_i^{(\alpha)}-x_j^{(\alpha)})^2
\prod_{\alpha<\beta, i,j}d(x_i^{(\alpha)}-x_j^{(\beta)}),
\end{equation}
where $d(x)=(L/\pi)\sin (\pi x/L)$
is the chord distance and $L$ is the
number of lattice sites.  Following techniques
developed for the SU(2) spin chain \cite{hala,shastry}
and  the OSp(2,1)
supersymmetric model \cite{kuramoto},
 it is straightforward to show that the above
SU($\nu$) Gutzwiller state without holes
gives the ground state for the SU($\nu$) spin chain
\cite{kawab,ha}.
The corresponding ground-state energy is thus computed as
\begin{equation}
E/L={\pi^2 \over 12} \big( {2-\nu \over \nu}
 + {1-2\nu \over  L^{2}} \big),
\end{equation}
which will be shown to coincide exactly with  the
result deduced from the ABA solution.

\vskip 7mm
\noindent
{\it 3.2. ABA solution}
\vskip 2mm

We now derive the ABA solution to the
SU($\nu$) spin chain, following the calculation
outlined in ref. \cite{kawab}. Let us begin with the
two-body scattering. In the asymptotic region
in the coordinate space, the two-body scattering
matrix for the above model is known to take the
simple form,
\begin{equation}
S_{ij}^{\alpha \beta}=  \lim_{\eta \rightarrow 0}
{k_i-k_j   +i \eta P_{\alpha \beta}
   \over k_i-k_j -i \eta  },
\end{equation}
in terms  the permutation operator
$P_{\alpha \beta}$ which interchanges the coordinates
$x_{\alpha}$ and  $x_{\beta}$.
Internal SU($\nu$) symmetry is now simply
taken into account via the operator $P_{\alpha\beta}$.
It is quite remarkable that this form of the $S$-matrix
is essentially the same as that for noninteracting electrons.
It should be noted, however, that this model actually
describes a non-trivial spin system  with long-range interaction.
We note that $S_{ij}^{\alpha \beta}$ satisfies the
Yang-Baxter factorization equation,
$S_{jk}^{\alpha \beta}S_{ik}^{\beta \gamma}
S_{ij}^{\alpha \beta}=S_{ij}^{\beta \gamma}
S_{ik}^{\alpha \beta}S_{jk}^{\beta \gamma}$.

Consider now the many-body scattering among particles
with $\nu-1$  different spins by taking  the $\nu$-th
species as the background. As mentioned above all
the particles obey the same statistics for  the
present SU($\nu$)  spin chain.
Imposing periodic boundary conditions
for the ring system with $L$ sites, we have to
diagonalize the scattering problem
\begin{equation}
e^{ik_jL}\Psi= S_{(j+1)j}  S_{(j+2)j}
 \cdots S_{(j-1)j} \Psi,
\end{equation}
where $\tilde S_{ij}=S_{ij}^{ij}$ \cite{yang}.
Introducing   $\nu-1$ kinds of
rapidities $k_j^{(\alpha)}$ for $\alpha=1,2, \cdots, \nu-1$,
one can  solve  this problem by the
nested BA method \cite{yang,suthc}.  We finally arrive at
the nested ABA equations for  the rapidities \cite{kawab},
\begin{equation}
 k_j^{(1)} L=  2\pi I_j^{(1)}+
\sum_{m} \Phi(k_m^{(2)}-k_j^{(1)})
+\sum_{l}\Phi(k_j^{(1)}-k_l^{(1)}),
\end{equation}
\begin{equation}
\sum_{l}\Phi(k_m^{(2)}-k_l^{(2)}) +2\pi I_{m}^{(2)}
=  \sum_{j} \Phi(k_m^{(2)}-k_j^{(1)})
+\sum_s \Phi(k_m^{(2)}-k_s^{(3)}),
\end{equation}
\centerline{$\cdots$}
\begin{equation}
\sum_{l}\Phi(k_s^{(\nu-1)}-k_l^{(\nu-1)})
+2\pi I_s^{(\nu-1)}
=  \sum_{j} \Phi(k_s^{(\nu-1)}-k_j^{(\nu-2)}),
\end{equation}
with $\Phi(k)= \pi {\rm sgn} (k)$,
where $I_j^{(\alpha)}$
classifies $\nu-1$ kinds of spin excitations,
\begin{equation}
I_j^{(\alpha)}={1 \over 2}(M_{\nu-1} + M_{\nu} + M_{\nu+1})
 \hskip 3mm {\rm mod} \hskip2mm 1,
\end{equation}
with $M_0=M_\nu=0$.  Here have introduced the quantity
\begin{equation}
M_\alpha=\sum_{\beta=\alpha}^{\nu-1} N_\beta,
\end{equation}
in terms of the number  of particles $N_\alpha$
with spin $\alpha=1, 2, \cdots, \nu$.
Henceforth $\nu-1$ kinds of spin
excitations are referred to as spinons
for simplicity. The energy is given in a
non-interacting form,
\begin{equation}
E=\sum_{j=1}^{M_1}
{1 \over 4}[(k_j^{(1)})^2-\pi^2] + C_\epsilon
\end{equation}
with the energy shift
\begin{equation}
C_\epsilon= {\pi^2 \over 12}(1-{1 \over L^2}) (L-2M_1) +
 {\pi^2 \over 6} (1-{1 \over L^2}).
\end{equation}

\vskip 7mm
\noindent
{\it 3.3. Bulk properties}
\vskip 2mm

We now introduce the density function   $\rho_\alpha (k)$
for the spin rapidity $k_j^{(\alpha)}$ in
the thermodynamic limit \cite{kawab}.  Note that
the density function  $\rho_\alpha (k)$ has constant values
in the region $[R_\alpha: B_{\alpha+1}<|k|<B_{\alpha}]$
with the condition  $B_1 \geq B_2 \geq,\cdots,
\geq B_{\nu-1}$  and $B_\nu=0$. These
density functions are easily evaluated as,
\begin{eqnarray}
 \rho_\alpha(k)=
\left\{
\begin{array}{rl}
{1 \over 2\pi}(\beta-\alpha+1)/ (\beta+1), & \hskip 5mm
\beta \geq \alpha, \\
0, & \hskip 5mm \beta < \alpha
\end{array}\right.
\end{eqnarray}
for the region $R_\beta$.
We note that $\rho_\alpha$ has a discontinuity
at every boundary $B_\alpha$.
The density of particles with spin $\alpha$ is
expressed in terms of $\rho_\alpha(k)$ as
\begin{equation}
n_\alpha= N_\alpha/L= \int_{-B_{\alpha}}^{B_{\alpha}}
 \rho_{\alpha}(k) dk
-\int_{-B_{\alpha+1}}^{B_{\alpha+1}}\rho_{\alpha+1} (k) dk
\end{equation}
for $\alpha=1,2, \cdots, \nu-1$ and
that for the $\nu$-th spin
is determined by the sum rule:  $\sum_{\alpha=1}^\nu n_\alpha=1$.
The cut-off parameter $B_\alpha$ is controlled by the
applied magnetic field through the relation,
\begin{equation}
B_\alpha=[\pi^2-2\alpha(\alpha+1)H]^{1/2},
\hskip5mm \alpha=1,2, \cdots, \nu-1,
\end{equation}
if  we assume that $\nu$ kinds of states have
magnetic moment $S, S-1, \cdots, -S$
with $\nu=2S+1$. Note that  $B_\alpha=\pi$
corresponds to the SU($\nu$) singlet state while $B_\alpha=0$
to the fully polarized state.

It is now straightforward to compute  bulk
static quantities \cite{kawab}. For example, the energy for
SU($\nu$) singlet is given  as
\begin{equation}
E/L={\pi^2(2-\nu) \over 12\nu},
\end{equation}
which agrees with the result (27)  obtained by
the SU($\nu$) Gutzwiller wavefunction.
Also, the magnetization is evaluated  as a
function of magnetic fields,
\begin{equation}
s_z  ={\nu-1 \over 2}- {1 \over 2\pi} \sum_{\alpha=1}^{\nu-1}
[\pi^2-2\alpha(\alpha+1)H]^{1/2} \theta(H_c^{(\alpha)}-H),
\end{equation}
where $\theta(x)$ is a step function and we introduced the
critical field  by $H_c^{(\alpha)}
=\pi^2/[2\alpha(\alpha+1)]$.
This yields the uniform spin susceptibility at low fields,
\begin{equation}
\chi_s={1 \over 2\pi^2}  \sum_{\alpha=1}^{\nu-1}
\alpha(\alpha+1).
\end{equation}
It is seen from (41) that the contribution to the
magnetization from $\alpha$-th spinons is saturated
at $H=H_c^{(\alpha)}$, and
the effective degrees of freedom for spinons are reduced
successively as the magnetic field is further increased.
Therefore spin-liquid
phases realized in  magnetic fields
are classified into $\nu-1$ sectors for which
each phase boundary  is determined by $H_c^{(\alpha)}$.
It is also easy to evaluate the velocity of each spinon excitation,
which is given  by the formula,
\begin{equation}
v_\alpha= {1 \over 2}[\pi^2-2\alpha(\alpha+1)H]^{1/2}.
\end{equation}
This velocity determines the specific-heat coefficient,
\begin{equation}
C/T ={\pi \over 3}
 \sum_{\alpha=1}^{\nu-1}{1 \over  v_\alpha}
\theta(H_c^{(\alpha)}-H).
\end{equation}
According to the finite-size scaling in CFT \cite{bcn},
the above expression for $C/T$ implies that $\nu-1$
kinds of spinon excitations are
described by independent $c=1$ CFT.
This point will be discussed in  detail below.

\vskip 7mm
\noindent
{\it 3.4.   Conformal properties}
\vskip 2mm

We now classify low-energy spinon excitations
in order to observe conformal properties \cite{kawab}.
Introduce first two
vectors $\vec m$ and $\vec d$ out of quantum numbers
associated with $\nu-1$ kinds of spinon excitations.
A quantum number $m_\alpha$ ($\alpha=1,2,\cdots,\nu-1)$
labels the change of the number of $\alpha$-th  spinons, i.e.
$m_\alpha=\Delta M_\alpha$. On the other hand,
$d_\alpha$ denotes an excitation which carries a momentum
$2\pi M_\alpha/N$. Hence the total momentum
is expressed as,
\begin{equation}
\Delta K= \sum_{\alpha=1}^{\nu-1} 2(\pi-\alpha k_F)d_\alpha,
\end{equation}
where $ k_F=\pi/\nu$ for $H=0$, and $1/L$ corrections
to the momentum transfer have been neglected.
According to boson nature of spinon
excitations,  the quantum numbers
$d_\alpha$ satisfy the selection rule
\begin{equation}
d_\alpha={1 \over 2}(m_{\alpha-1}+m_{\alpha+1})
\hskip5mm {\rm mod} \hskip 3mm 1,
\end{equation}
for $  1 \leq  \alpha \leq \nu-1 $ with the condition
$m_0=m_\nu=0$,  which directly follows from
eq. (33).  Low-energy excitations are now
classified in the matrix formula,
\begin{equation}
\Delta E= {2 \pi \over L} \sum_{\alpha=1}^{\nu-1}
v_\alpha x_\alpha,
\end{equation}
where $x_\alpha$ is the scaling dimension for
each spinon with spin $\alpha$,
\begin{equation}
x_\alpha={1 \over 4} ({\bf Z}^{-1} \vec m)_\alpha^2
+({\bf Z}^{t} \vec d)_\alpha^2 +n_\alpha^+ +n_\alpha^- ,
\end{equation}
and $n_\alpha^\pm$ labels  particle-hole type
excitations for the  $\alpha$-th spinons.
The $(\nu-1) \times (\nu-1)$ matrix $\bf{Z}$, which is
referred to as the {\it dressed charge matrix}
\cite{izergin}, is given as \cite{kawab}
\begin{eqnarray}
Z_{\alpha\beta}=\left\{
\begin{array}{rl}
\alpha [\beta(\beta+1)]^{-1/2},  & \hskip 3mm
\alpha \leq \beta \leq \nu-1, \\
0, & \hskip 3mm \alpha > \beta .
\end{array}\right.
\end{eqnarray}
One can see  that
the above scaling dimension has the typical form inherent in $c=1$
CFT \cite{review} and that non-negative integers
$n_\alpha^\pm$ form conformal towers which characterize the
representation of Virasoro algebra \cite{izergin}.

Let us now consider  the total scaling
dimension \cite{kawab},
\begin{equation}
x=\sum_{\alpha=1}^{\nu-1}x_\alpha,
\end{equation}
which determines the critical exponents  $\eta=2x$
for correlation functions.  It turns out that $x$ is expressed
in terms of the Cartan matrix ${\bf C}$
for SU($\nu$) Lie algebra,
\begin{equation}
x={1 \over 4} \vec m^t {\bf C} \vec m
+ \vec d^t {\bf C}^{-1} \vec d +
\sum_{\alpha=1}^{\nu-1} (n_\alpha^+ +n_\alpha^-),
\end{equation}
where
\begin{equation}
{\bf C}=
\left(
\matrix {2  & -1      & \null  & \null   \cr
         -1         & 2       &  -1    & \null   \cr
        \null       & \ddots  & \ddots & \ddots  \cr
        \null       & \null   &  -1    & 2       \cr}
\right) .
\end{equation}
A remarkable point is that this expression for $x$
holds {\it  even in
magnetic fields} although there does not exist SU($\nu$) symmetry
and the velocities of spinons are strongly
modified.  These characteristic properties are in contrast to
those for the SU($\nu$) antiferromagnetic Heisenberg chain
with nearest neighbor interaction
for which the scaling dimension changes continuously
and the Cartan matrix can specify excitations
only at $H=0$ \cite{suzuki}.
Therefore the correlation exponents are not affected by
magnetic fields in the present model,
and are determined   by those of $c=\nu-1$
CFT.

\vskip 7mm
\noindent
{\it 3.5.  Correlation exponents}
\vskip 2mm

Following standard methods in  CFT
\cite{bpz,bcn,cardy}, we deduce all the critical exponents of
various correlation functions \cite{kawab}.
We apply the assignment of quantum numbers
to a given correlation function, which was
used for the Hubbard model \cite{fk}
and subsequently applied to
the {\it t-J} model with nearest-neighbor hopping
\cite{kyc}.  For example, let us  consider
the long-distance behavior of the spin
correlation functions characterized
by the exponent $\beta_j$,
\begin{equation}
<S_z(r)S_z(0)>  \simeq  \sum_{j=1}^{\nu-1}
A_j \cos(2jk_F r) r^{-\beta_j}.
\end{equation}
This correlation function  conserves the
number of particles,  so that we set $\vec m=(0,\cdots,0)$.
There remain several choices for the quantum numbers
$\vec d$.  A choice of  $d_\alpha=\delta_{\alpha j}$
($1 \leq  j \leq \nu -1$) results in the $2jk_F$
oscillation piece of the  spin correlation functions.
We thus obtain spin-correlation
exponents for the $2jk_F$ oscillation part
\cite{kawab},
\begin{equation}
\beta_{j}={2j(\nu-j) \over \nu},
\hskip 5mm  1 \leq  j \leq \nu-1.
\end{equation}
For the SU(2) case, this result agrees with  that
of the Gutzwiller wavefunction
at half filling \cite{vol}.
We stress here again that the above critical
exponents do not depend on magnetic fields.

\vskip 15mm
\noindent
\begin{center}
\begin{bf}
 \S 4. OSp($\nu,1$) Supersymmetric {\it t-J} Model
\end{bf}
\end{center}
\vskip 3mm

Now we wish to observe what is modified
  when we dope
holes into the SU($\nu$) Haldane-Shastry model.  This problem
seems interesting since  highly correlated electron systems
have attracted particular attention recently.  Such an attempt
was firstly done for the OSp(2,1) supersymmetric model
\cite{kuramoto}, and
subsequently the
model was extended to more generic supersymmetric model with
OSp($\nu$,1) symmetry \cite{kawab,ha}.
 We give a brief review of ref. \cite{kawab}
here.

\vskip 7mm
\noindent
{\it 4.1. Symmetry properties }
\vskip 2mm

We first note that if the fermion numbers are taken as
$F(\beta)=(0, \cdots, 0, 1)$
for $\beta =(1,2,\cdots, \nu+1)$ in the Hamiltonian
(23), this model is equivalent to the
multicomponent {\it t-J} model \cite{kawab,ha},
\begin{equation}
{\cal H}= - \sum_{i<j} \sum_{\alpha=1}^{\nu}
t_{ij} c_{i\alpha}^{\dag} c_{j\alpha}+
\sum_{i<j} {J_{ij} \over 2} [\sum_{\alpha , \beta \leq \nu}
X_i^{\alpha\beta}X_j^{\beta\alpha}- (1-n_i)(1-n_j)],
\end{equation}
with the so-called {\it supersymmetric condition}
$t_{ij}=J_{ij}$, where $n_i$ is the electron
number at the $i$-th site.
Here the double occupation of every site is
strictly prohibited.
We have used the fact that
the term like $X_i^{\alpha(\nu+1)}X_j^{(\nu+1)\alpha}$  with
$\alpha=1,2,\cdots, \nu$ corresponds to the hopping term
of electrons with  spin $\alpha$.
For a special case with three-component particles
with $F=(0,0,1)$, the model reduces
to the OSp(2,1) supersymmetric {\it t-J}  model
(apart from the chemical potential term) \cite{kuramoto},
\begin{equation}
{\cal H}= - \sum_{i<j, \sigma}t_{ij}
c_{i\sigma}^{\dag} c_{j\sigma}+
\sum_{i<j}J_{ij}[ {\bf S_i \cdot  S_j} - {1 \over 4}n_in_j],
\end{equation}
which is obtained by
doping holes into the SU($2$) Haldane-Shastry model.

Here, we briefly mention  supersymmetry properties of the
above {\it t-J} model.  For the graded model (23) or (55)
with bosons and fermions, it is known that
the Hubbard operator should satisfy the following
commutation relation \cite{wiegmann,forster},
\begin{equation}
[X_i^{\alpha\beta}, X_j^{\gamma\eta}]_\pm
=\delta_{ij}(\delta_{\beta\gamma}X_i^{\alpha\eta}
\pm \delta_{\alpha \eta} X_i^{\gamma \beta}),
\end{equation}
where the anti-commutator (+) should be used only for the fermion
operators.  This algebra is called as
the doubly-graded  Lie superalgebra
with Osp($\nu,1$) supersymmetry.
Therefore the model (55) is invariant under the global OSp($\nu,1$)
transformation \cite{wiegmann,forster}.
Thanks to the supersymmetry
of the Hamiltonian, we can successfully
apply the ABA method to the above {\it t-J} model \cite{kawa}.

\vskip 7mm
\noindent
{\it 4.2. Ground-state wavefunction}
\vskip 2mm

Before proceeding with the ABA calculation, we
mention the ground-state wavefunction.
The ground state of the OSp($\nu,1$) model
is given by the SU($\nu$) Gutzwiller
state {\it with holes}, as was
demonstrated for OSp(2,1) case \cite{kuramoto}.
In this case the Gutzwiller projection
operator in (24) projects out configurations with more than
one electron on each lattice site, which
implies that there can be cites without any electrons
(assigned as holes). Taking the fully polarized state $|F>$
as the reference state, the
Gutzwiller state with holes is written as
\cite{kuramoto,kawab,ha},
\begin{equation}
 | \Psi_G>= \sum_{(\alpha,i),j}
\Psi_G(\{x_i^{(\alpha)}\}, \{s_j\})
\prod_{\alpha, i} b_{i}^{(\alpha) \dag}
\prod_{j} h_{j}^{\dag}|F>,
\end{equation}
where $b_{i}^{(\alpha) \dag}$ is
a creation operator for spin particles
and $h_{j}^{\dag}=a_{j}^{(\nu)}$ is
that for doped holes. The Gutzwiller
wavefunction is now expressed in terms of coordinates
of spin particles ($x_i^{(\alpha)}$) and holes ($s_m$) as,
\begin{eqnarray}
\Psi_G(\{x_i^{(\alpha)}\}, \{s_j\})
=  \exp[-i\pi(\sum_{\alpha,i}x_i^{(\alpha)}+\sum_{j}s_j)]
  \prod_{\alpha, i<j} d(x_i^{(\alpha)}-x_j^{(\alpha)})^2
\nonumber \\
 \times \prod_{\alpha<\beta, i,j}d(x_i^{(\alpha)}-x_j^{(\beta)})
\prod_{\alpha, i,m} d(x_i^{(\alpha)}-s_m)
\prod_{m<n} d(s_m-s_n)
\end{eqnarray}
One can show easily that this gives the ground-sate
wavefunction of the OSp($\nu,1$) supersymmetric
{\it t-J} model, the detail for which can be found in refs.
\cite{kuramoto,kawab,ha}.

\vskip 7mm
\noindent
{\it 4.3. ABA solution}
\vskip 2mm

We shortly outline how to deduce the ABA solution to the
Osp($\nu,1$) supersymmetric {\it t-J} model \cite{kawab}.
A remarkable point is that
the two-body $S$-matrix for this model is given by the
same formula (28) as in the SU($\nu$) spin chain, and
symmetry properties can be taken into account
solely via the permutation operator $P_{\alpha\beta}$.
Hence, using the $S$-matrix (28) for the two-body scattering, it
is now straightforward  to deduce the ABA solution to the
OSp($\nu,1$) {\it t-J} model.  Taking  the reference
state full of particles with the $\nu$-th spin,
consider now the scattering problem among
$\nu-1$ kinds of spinons and holons (holes).
In order to diagonalize this problem, therefore,
besides $\nu-1$ kinds of
spin rapidities $k_j^{(\alpha)}$ ($\alpha =1,2,
\cdots, \nu-1)$, it is necessary to introduce the charge rapidity
$k_j^{(\nu)}$ \cite{suthc}. As a result
the last line of the ABA equations in (32)
should be modified by grading holes \cite{kawab},
\begin{equation}
 k_j^{(1)} L=  2\pi I_j^{(1)}+
\sum_{m} \Phi(k_m^{(2)}-k_j^{(1)})
+\sum_{l}\Phi(k_j^{(1)}-k_l^{(1)}),
\end{equation}
\begin{equation}
\sum_{l}\Phi(k_m^{(2)}-k_l^{(2)}) +2\pi I_{m}^{(2)}
=  \sum_{j} \Phi(k_m^{(2)}-k_j^{(1)})
+\sum_s \Phi(k_m^{(2)}-k_s^{(3)}),
\end{equation}
\centerline{$\cdots$}
\begin{equation}
2\pi I_s^{(\nu)}
=  \sum_{j} \Phi(k_s^{(\nu)}-k_j^{(\nu-1)}), \hskip5mm
s=1, 2, \cdots,  M_{\nu},
\end{equation}
where $I_s^{(\nu)}$  is a quantum number
which labels the degrees of freedom for doped holes.
According to the antisymmetric nature of
electrons, the selection rule for this quantum number reads,
\begin{equation}
I_s^{(\nu)}={1 \over 2} M_\nu \hskip 5mm  {\rm mod} \hskip 2mm 1.
\end{equation}

\vskip 10mm
\noindent
{\it 4.4. Bulk properties}
\vskip 2mm

We calculate bulk quantities
following techniques outlined for the spin chain.
It is remarkable that  the shape of the
the calculated  magnetization curve does not depend on
electron concentrations, which is given by  the
same formula as (41), but only the
critical field
\begin{equation}
H_c^{(\alpha)}
={\pi^2n(n-2) \over 2\alpha(\alpha+1)}
\end{equation}
is modified by hole-doping ($n$: electron density) \cite{kawab}.
Also, other bulk quantities are easily calculated.
By the second derivative of the ground state energy
with respect to $n$, we obtain
the charge susceptibility (compressibility)
as a function of electron concentrations,
\begin{equation}
\chi_c={2\nu \over \pi^2(1-n)},
\end{equation}
which shows a divergent behavior near
 insulating phase $n=1$ \cite{kuramoto,kawab,ha}.
The coefficient of the  $T$-linear
heat capacity is given
in terms of the velocities of spinons and
holons \cite{kuramoto,kawab,ha},
\begin{equation}
\gamma ={\pi \over 3}[
{1 \over v_c} +
 \sum_{\alpha=1}^{\nu-1}{1 \over  v_\alpha}
\theta(H_c^{(\alpha)}-H)],
\end{equation}
where the velocity of spinon excitations, $v_\alpha$,
takes  the same formula as
in (43) while that for holon excitations depends
on the electron concentration,
\begin{equation}
 v_c= \pi(1-n)/2.
\end{equation}
The above expression for the specific heat implies that
the critical behavior of the
present model is described by $c=1$ CFT.
We can see that there are the Luttinger-liquid relation among
bulk quantities, $ \pi \chi_c v_c= \nu$.

\vskip 7mm
\noindent
{\it 4.5. Conformal properties}
\vskip 2mm

We now classify  low-energy spin and charge excitations
in order to study conformal properties \cite{kawab}.
Let  us introduce $\nu$-component vectors $\vec m$ and  $\vec d$
for  quantum numbers which specify  excitations.
We assume  the $\nu$-th component of
vectors to be related to  the holon degrees of freedom.
According to the antisymmetry properties of electron
wavefunction, the selection rule for quantum numbers  reads
\begin{equation}
d_\alpha={1 \over 2}(m_{\alpha-1}+m_{\alpha+1})
\hskip 5mm {\rm mod} \hskip 3mm 1,
\end{equation}
for spinon excitations ($\alpha=1, 2, \cdots, \nu-1$), whereas
that for holon excitations is
\begin{equation}
d_\nu={1 \over 2}(m_{\nu-1}+m_{\nu})
\hskip 5mm {\rm mod} \hskip 3mm 1.
\end{equation}
Note that quantum numbers $d_\alpha$ carry the
large momentum transfer $2(\pi-\alpha k_F)d_\alpha$ for
$1 \leq \alpha \leq \nu$ with Fermi momentum
$k_F=\pi n/ \nu$.   The excitation spectrum is now
classified  as
\begin{equation}
\Delta E= {2 \pi \over L} \sum_{\alpha=1}^{\nu} v_\alpha
x_\alpha,
\end{equation}
from which we can read the scaling
dimension  $x_\alpha$. The resulting scaling dimension
is expressed in the same formula as (48) by
extending the dressed charge matrix  \cite{izergin} to
the  $\nu \times \nu$ matrix  $\bf{Z}$,
\begin{eqnarray}
Z_{\alpha\beta}= \left\{
\begin{array}{rl}
\alpha [\beta(\beta+1)]^{-1/2},  & \hskip 3mm
\alpha \leq \beta \leq \nu-1, \\
\alpha /\sqrt{\nu}, & \hskip3mm \beta=\nu, \hskip 2mm
 1 \leq \alpha \leq \nu \\
\end{array}\right.
\end{eqnarray}
and $Z_{\alpha\beta}=0$ otherwise \cite{kawab}. Matrix elements of
$\nu$-th row and the $\nu$-th column are related to the holon
degrees of freedom.
{}From the expression (70), one can see that
that holon excitations as well as spinon excitations
are described by independent $c=1$ CFT.
Namely this liquid is classified as the Luttinger
liquid.  The total scaling dimension
$x=\sum_{\alpha=1}^{\nu}x_\alpha$
is reduced to the simple expression  (50)
where the $\nu \times \nu$ matrix $\bf{C}$ in this case is
\begin{equation}
{\bf C}=
\left(
\matrix {2  & -1      & \null  & \null   \cr
         -1         & \ddots       &  -1    & \null   \cr
        \null       & \ddots  & 2 & \ddots  \cr
        \null       & \null   &  -1    & 1       \cr}
\right)
\end{equation}
which is nothing but the Cartan matrix for the OSp($\nu,1$)
Lie superalgebra \cite{kawab}.  We wish to emphasize
here again that the above scaling dimension depends
neither on  magnetic fields nor on electron concentrations,
characterizing the universality class of
quantum $1/r^2$ models.

\vskip 7mm
\noindent
{\it 4.6.  Correlation exponents}
\vskip 2mm

Let us now evaluate critical  exponents for
various correlation functions with the
use of  CFT.
First we compute the critical exponents of the
spin correlation function.
By choosing the quantum numbers
 $d_\alpha=\delta_{\alpha j}$
($1 \leq  j \leq \nu-1$) for the spin correlation
function \cite{fk,kyc,kawa,fs}, we obtain the
spin correlation exponents for the $2jk_F$
oscillation part,
\begin{equation}
\beta_{j}=2j, \hskip 5mm  1 \leq  j \leq \nu-1.
\end{equation}
Remarkably enough, all the exponents are given by
the canonical (integer) values.
Also, critical
exponents for the charge correlation functions are obtained
similarly, which are given by the same formula as for the spin
exponents, but the $2\nu k_F$ oscillation piece appears with
$\beta_\nu=2\nu$. It is instructive
to note  that the spin correlation exponents
discontinuously change when the holes are doped into the
spin chain.  This can be checked by comparing the expressions
(54) and (73).  This property of the discontinuity
is known to be common in the correlated electron systems
close to the insulator.

We now evaluate the correlation exponent $\eta$ of
the field correlator of electrons,
$<c^{\dag}_{\alpha}(r)c_{\alpha}(0)>\simeq \cos (k_F r)
r^{- \eta }$. A simple excitation relevant to
this correlator is given by
the set of quantum numbers $(m_1,m_2,\cdots,m_\nu)
=(1,1,\cdots,1)$ and
$(d_1,d_2,\cdots,d_\nu)=(1/2,0,\cdots,0)$ \cite{fk,kyc,kawa,fs}.
The momentum carried by this excitation is
$\pi(1-k_F)$.  The resultant critical exponent turns out
to be  $\eta=1$ \cite{kawab}.  Fourier transformation of
this correlation functions gives
the momentum distribution
\begin{equation}
n_k= n_{k_F} -{\rm const.}
|k-k_F|^{\theta} {\rm sgn}(k-k_F),
\end{equation}
with the corresponding exponent $\theta=\eta-1=0$.
By taking account the fact that there is no
logarithmic correction in the present case,
$\theta=0$ implies that there should be
a discontinuity in the momentum distribution
at the Fermi point, as  firstly
pointed out for the OSp($2,1$) model \cite{kuramoto}. Recall again
that all the critical exponents of correlation functions
are given by the {\it canonical} values  as well known for the
Gutzwiller wavefunction \cite{vol}.
At first glance these canonical
exponents  seem  to be contradicted  to Luttinger
liquid theory \cite{halc}, but we should keep in
mind that this fixed point is indeed
on the critical line of the Luttinger
liquid ($c=1$ Gaussian CFT).
The present model is, therefore, classified as a specific
example of the Luttinger liquid which has canonical
exponents \cite{kuramoto}.

We have not been concerned here with
thermodynamic properties at arbitrary temperatures.
In order to describe the full thermodynamics
of the {\it lattice} models, it
is crucial to study the degeneracy of  excited
states.  Concerning this problem, the
free-spinon picture of Haldane \cite{halc},
Yangian symmetry \cite{halz}
 and related methods \cite{wang,ssb}
have been successfully used to construct
correct thermodynamics.

\vskip 15mm
\begin{center}
\begin{bf}
\S 5.  Hierarchical Models Related to FQHE
\end{bf}
\end{center}
\vskip 3mm

One of the most interesting aspects of the $1/r^2$
quantum systems is the intimate relationship
\cite{halc,kawac,poly,for,kawad,iso}
to the fractional quantum Hall effect (FQHE)
\cite{lag,hie,jain}.
 We have indeed seen that
the construction of eigenstates for  the $1/r^2$ models
is quite analogous to that for the FQHE:
the ground state is given by the Jastrow
wavefunction, and the excited states are constructed by multiplying
polynomials to the ground state wavefunction.
Therefore, besides much interest in the integrability,
these quantum models should exhibit interesting phenomena
related to the FQHE.
We have recently proposed a novel hierarchy of the 1D
quantum models with $1/r^2$ interaction \cite{kawac},
the construction of which is
essentially same as that of a hierarchical FQHE
with the filling fraction \cite{hie,jain},
\begin{equation}
 f_\nu =
{1 \over \displaystyle p_1-
{1 \over \displaystyle p_2
{\cdots_{\displaystyle -
{1 \over \displaystyle p_\nu}}}}}.
\end{equation}
In particular the matrix deduced from the energy spectrum
has been shown to coincide with  the {\it topological-order
matrix} which characterizes the internal
structure of the FQHE state
\cite{wen,blok,zee}. In this section we  briefly review
the hierarchical models related to the FQHE \cite{kawac,kawad}.

\vskip 8mm
\noindent
{\it 5.1. Continuum  Models and ABA solution}

\noindent
Let us introduce a family of $\nu$-component electron
models ($\alpha=1,2, \cdots, \nu$)
with  $1/r^2$ interaction
in the periodic ring of length $L$ \cite{kawac},
\begin{equation}
 H= -{1 \over 2}\sum_{i}
{\partial^2 \over \partial x_i^2}
+\sum_{\alpha \leq \beta}\sum_{i<j} d_{ij}^{-2}
V_{\alpha\beta}(V_{\alpha\beta}
+P_{ij}^{\alpha\beta}),
\end{equation}
with chord distance $d_{ij}=(L/\pi)\sin[\pi(x_i-x_j)/L]$,
where  $P_{ij}^{\alpha\beta}$
is the spin exchange operator with spin indices
$\alpha, \beta=1,2, \cdots \nu$, and
$V_{\alpha\beta}$ is the interaction parameters
dependent on spin indices.
Starting from noninteracting SU($\nu$)
electrons (zero-th stage),  let us specify
the stage of the hierarchy in terms of the matrix
formula of the interaction parameter.
The first  stage is defined  by introducing
interaction $\lambda_1$ uniformly to all the electrons,
i.e.  $V_{\alpha\beta}=\lambda_1$.
The resultant model is the SU$(\nu)$
Sutherland model for which
only the charge excitation is affected by the
interaction \cite{ha,polya,hikami}.
The second family is introduced by
turning on the uniform interaction $V_{\alpha\beta}=\lambda_2$
among electrons except for the $\alpha=1$ species.
By iterating this procedure step by step, the $\nu$-th
stage of the model is characterized by the
interaction matrix \cite{kawac},
\begin{equation}
{\bf  V}=
\left(\matrix {\lambda_1  & \cdots      &  & \lambda_1    \cr
             \vdots      & \ddots     &     & \vdots     \cr
                 &       &                &    \cr
              \lambda_1   &   \cdots    &        & \lambda_1 \cr}
\right)
+
\left(\matrix {0  & \cdots      &        & 0    \cr
             \vdots    & \lambda_2  & \cdots      & \lambda_2  \cr
                   &    \vdots   & \ddots   &  \vdots  \cr
              0   &   \lambda_2    & \cdots       & \lambda_2 \cr}
\right)
+ \cdots +
\left(\matrix {0  & \cdots      &        & 0    \cr
             \vdots    & \ddots    &     & \vdots     \cr
                &       &   0 &  0  \cr
              0   &  \cdots     &   0     & \lambda_\nu \cr}
\right) .
\end{equation}
Note that the interaction $\lambda_m$
acts on particles with spin indices
$\alpha=m, m+1, \cdots, \nu$, hence
it is regarded as  spin-dependent interaction.
The construction of the above hierarchy is
quite analogous to that for the FQHE \cite{hie,jain},
in particular, to Jain' composite-fermion
construction of the hierarchical FQHE
\cite{jain}.
In the FQHE case, the interaction matrix
introduced here serves as a topological order matrix
which specifies the internal structure of the FQHE
\cite{wen,blok,zee}.
We shall see that the interaction matrix
completely specifies the low energy excitations,
and that the present model can describe
essential properties for edge states of the
hierarchical FQHE with filling fraction $f_\nu$
\cite{wen}.

Let us now find the solution to the problem
by the ABA method \cite{kawac}.
We first consider the scattering of two particles
by the interaction $d_{ij}^{-2}V_{\alpha\beta}
(V_{\alpha\beta} + 1)$
As mentioned in the previous sections,
this interaction yields the $S$-matrix of
$\exp [-i\phi_{\alpha\beta}(k_i-k_j)]$
with the phase shift function,
\begin{equation}
\phi_{\alpha\beta}(k)=V_{\alpha\beta}\pi{\rm sgn}(k),
\end{equation}
in the asymptotic region ($|x_i-x_j| >>1$).
The step-like form of the phase shift $\phi_{\alpha\beta}(k)$ is
characteristic of  the $1/r^2$ interaction.
Taking into account the spin degrees of freedom,
we then find the two-body $S$-matrix for
the above hierarchical models as \cite{kawac},
\begin{equation}
S_{ij}= \lim_{\epsilon \rightarrow 0}
{k_i-k_j  -i \epsilon P_{ij}^{\alpha\beta} \over
k_i-k_j -i \epsilon}  \
e^{-i\phi_{\alpha\beta}(k_i-k_j)},
\end{equation}
where the first factor arises from non-interacting
SU($\nu$) electrons, while the  second
is due to the $1/r^2$ interaction.
The key feature that  the $S$-matrix (79) is
a simple superposition of two $S$-matrices
makes it possible to treat
the scattering problem consistently by
nested Bethe ansatz techniques.

As mentioned in the previous section,
the essence of the  ABA method is that
the many-body $S$-matrix in this family can be
decomposed into two-body matrices
in spite of  long-range nature of interaction.
For a simple case of SU($\nu$) symmetry,
we have already checked that this
solution gives the exact spectrum \cite{kawab,kawad}.
If this is the case in general,
we can diagonalize the many body $S$-matrix
by the standard nested Bethe ansatz.
Consequently we arrive at the  ABA equations for
$\nu$-kinds of rapidities $k_j^{(\alpha)}$ \cite{kawac},
\begin{equation}
 k_j^{(1)} L=  2\pi I_j^{(1)}+
\sum_{m=1}^{M_2} \Phi(k_m^{(2)}-k_j^{(1)})
+ \lambda_1 \sum_{l=1}^{M_1}\Phi(k_j^{(1)}-k_l^{(1)}),
\end{equation}
\begin{equation}
(\lambda_\alpha+1)\sum_{l=1}^{M_\alpha}
\Phi(k_m^{(\alpha)}-k_l^{(\alpha)})
+2\pi I_{m}^{(\alpha)}
= \sum_{s=\pm 1} \sum_{j=1}^{M_{\alpha+s}}
\Phi(k_m^{(\alpha)}-k_j^{(\alpha+s)}),
\end{equation}
for $2 \leq \alpha \leq \nu$, where $\Phi(k)=\pi {\rm sgn}(k)$
and $I_j^{(\alpha)}$ is an integer or a  half integer
which classifies the charge and spin excitations.
In the above equations we have introduced the quantity
$M_\alpha=\sum_{\beta=\alpha}^{\nu} N_\beta$
where $N_\beta$ is the number of electrons with spin $\beta$.
The energy is  written in the noninteracting form $E=(1/2)
\sum_{j}(k_j^{(1)})^2$.

\vskip 8mm
\noindent
{\it 5.2.  Bulk properties}
\vskip 2mm

Let us calculate bulk quantities \cite{kawac}.
It is  remarkable that all the bulk quantities
are expressed solely by the parameter
$f_\nu$ introduced in (75)
if we assign the set of parameters $p_i$
in terms of interaction parameters $\lambda_i$ as,
\begin{equation}
p_i=\lambda_i+2 -\delta_{i1},
\end{equation}
for $\alpha=1,2, \cdots, \nu$.
For instance,  the  ground-state
energy is evaluated as
\begin{equation}
E_g/L=\pi^2 n^3/(6f_\nu^2),
\end{equation}
where $n$ is the electron density.
The second  derivative of $E_g(n)$ yields
the charge susceptibility (compressibility)
 in terms of $f_\nu$,
\begin{equation}
\chi_c= (f_\nu/\pi)^2  n^{-1}.
\end{equation}
Roughly speaking, the quantity $1/f_\nu$ corresponds to
the size of the exchange-correlation hole
due to the $1/r^2$ interaction:
we can regard the effective
volume of electrons to become $1/f_\nu$ times
as large as that of free electrons.

The free energy at finite
temperatures ($T$) is formulated as,
\begin{equation}
F= -\mu -(T/2\pi) \int_{-\infty}^{\infty}
\log [1+ \exp(-\epsilon_1(k))] dk
\end{equation}
in terms of the dressed energy,
\begin{equation}
\epsilon_1(k)/T= {1 \over 2}k^2-\mu
-\log [1+ \exp(-\epsilon_2(k))]
+\lambda_1  \log [1+ \exp(-\epsilon_1(k)/T)] ,
\end{equation}
\begin{equation}
\epsilon_\alpha(k)/T= \sum_{s=-1,0,1} (-1)^s (1
+\lambda_\alpha \delta_{q0})
\log [1+ \exp(-\epsilon_{\alpha+s}(k))/T],
\end{equation}
for $\alpha=2,3, \cdots, \nu $, with
$\epsilon_{\nu+1}=\infty$. The low-temperature expansion
of the free energy yields the coefficient
of the $T$-linear specific heat as,
\begin{equation}
C/T = {\pi  \over 3 v}
\end{equation}
with the velocity $ v=\pi n/(2 f_\nu)$.
Note that all the velocities for $\nu$ kinds of
elementary excitations
have the same value $v$ although there is not SU($\nu$)
symmetry in the model.
The Luttinger-liquid relation between the
charge susceptibility and the velocity takes the form,
$\pi \chi_c v= f_\nu/2$, which can determine
the critical behavior of charge excitations.

\vskip 5mm
\noindent
{\it 5.3.  Excitations}

Let us now turn to the excitation spectrum \cite{kawac}.
Using the ABA equations,
low-energy excitations are classified
in the matrix formula,
\begin{equation}
\epsilon= {2 \pi v \over L}[{1 \over 4}\vec m^t {\bf T} \vec m
+   \vec d^t ({\bf T})^{-1} \vec d],
\end{equation}
where the $\nu \times \nu$ matrix ${\bf T}$ is
evaluated as,
\begin{equation}
{\bf T}=
\left(
\matrix {p_1  & -1      & \null  & \null   \cr
         -1         & p_2       &  \ddots    & \null   \cr
        \null       & \ddots  & \ddots & -1  \cr
        \null       & \null   &  -1    & p_\nu       \cr}
\right).
\end{equation}
Here  the $\nu$-component vector $\vec m$ is out of
quantum numbers which classify the charge  and spin
excitations,  where we assumed that $m_1$ and $d_1$
label charge excitations.
It is remarkable that the above matrix ${\bf T}$ deduced
from the energy spectrum is nothing but the topological-order
matrix for the FQHE with filling
fraction $f_\nu$ \cite{blok}.
One can see that there is $\nu$ free parameters, $p_i$, in the
above matrix.  This implies that the critical behavior of the
present model is described by the corrections of $ \nu$
independent CFT with the central charge $c=1$, where $p_i$ can make
$c=1$ critical lines.

\vskip 8mm
\noindent
 {\it 5.4.  t-J model and lattice effects}
\vskip 2mm

To see the relationship to the composite fermion theory
in the FQHE more explicitly, let us
observe what happens for the lattice case \cite{kawac}.
We introduce a hierarchical family of
{\it t-J} models with $1/r^2$ interaction,
\begin{equation}
{\cal H}=  \sum_{\alpha, i \neq j}
d_{ij}^{-2}c_{i\alpha}^{\dag} c_{j\alpha}
+\sum_{\alpha \leq \beta, i<j}
d_{ij}^{-2} V_\beta(V_\beta+P_{ij}^{\alpha\beta}),
\end{equation}
with interaction parameters  $V_{\alpha\beta}$
defined in (77). Here  configurations with more
than one electron at each site are assumed to be prohibited.
The first family of  the hierarchy coincides with
the SU($\nu$) {\it t-J} model introduced in \cite{ha}.
This class of lattice models can be solved
by the ABA method, and the resulting  nested ABA equations
are given by the same formula as in (80) and (81).
Hence,  the bulk quantities are given by the
same expressions as for the continuum case.
In the lattice  {\it t-J} model, however, we encounter two crucial
constraints \cite{kawac}:
(a) parameter $V_{\alpha\beta}$ should be an
even positive integer, and
(b) the rapidity $k_j^{(\alpha)}$ should be in the
region [$-\pi, \pi$] as is the case for ordinary lattice models.
According to (a), the parameter $f_\nu$  should be a
fraction with the {\it odd} denominator, which demonstrates
the analogy to FQHE  explicitly.  Furthermore,
the constraint (b) brings about a remarkable
property, i.e. a singular property
at the electron density $n=f_\nu$  above  which
the Luttinger-liquid state breaks down
($n > f_\nu$) \cite{kawac}.
This means  that a band-edge singularity
for non-interacting SU($\nu$) lattice electrons
at the density $n=\nu$ is modified into the singularity
at the fractional filling
 $n=f_\nu$ in the presence of the $1/r^2$ interaction.
We note that this phenomenon is essentially the same as that
for the composite fermion theory of the FQHE, i.e.
the hierarchical FQHE with
the filling $n=f_\nu$ can be realized by  starting from the
integer (noninteracting) QHE with the filling
$\nu$ \cite{jain}.

\vskip 8mm
\noindent
{\it 5.5. Chiral constraint and FQHE edge states}
\vskip 2mm

In order to observe how remarkably the present
model reproduces essential properties expected for
edge states of the FQHE, let us think of what will happen
if we consider only right
(or left)-moving electrons in the 1D system \cite{kawac,kawad}.
This is referred to as {\it chiral constraint} which
is essential for edge
states of the FQHE in disk geometry \cite{wen}.
In order to deal with  chiral constraint, we
make use of  a trick valid for the $1/r^2$ systems.
We first add two electrons
at the left and right Fermi points, in order to
suppress the current $\vec d$ which does not
exist in the chiral model.
This results in the energy increment both for the
right and left branches.  Dividing
the energy increase into two parts and discarding
the right-going piece, we obtain the excitation spectrum
for left-going piece as \cite{kawac,kawad},
\begin{equation}
\epsilon ={\pi v \over L} \vec m^t {\bf T} \vec m.
\end{equation}
This formula reproduces the
spectrum described by the holomorphic piece of
$\nu$ independent  $c=1$ CFT.  Hence we can
determine the critical exponents for various
correlation functions. For example,
the critical exponent $\theta_\alpha$
for the momentum distribution function
\begin{equation}
n_k^{(\alpha)} \simeq {\rm const.}
+ a_0 \, {\rm sgn}(k-k_F)|k-k_F|^{\theta_\alpha}
\end{equation}
is obtained  as
\begin{equation}
\theta_\alpha= \sum_{j=1}^{\alpha} \lambda_j.
\end{equation}
All the other exponents can be similarly obtained.
We stress here that all the critical
exponents  agree with those of
effective field theory (chiral Luttinger
liquids) for the edge states of FQHE
with the fraction $f_\nu$ \cite{wen}.
In particular  the key matrix ${\bf T} $
deduced from the excitation spectrum
coincides exactly with the  topological order matrix
which characterizes  the internal structure
of the  FQHE  with the filling fraction $f_\nu$
\cite{blok,zee}.

We have observed  that the hierarchical models considered
here describe characteristic  properties for the edge
states of the FQHE with filling fraction $f_\nu$
so remarkably. The correspondence is not accidental, and
we can indeed see a clear reason for it in  the role played by the
phase shift function  $\lambda_\alpha \pi {\rm sgn} (k)$.
Recall that in the composite fermion theory for the
FQHE \cite{jain},  flux quanta are attached
to electrons in order to evolve the hierarchy
starting from noninteracting electron systems \cite{jain,wen,blok,zee}.
A crucial point is that attaching  $\lambda_\alpha$ flux quanta
in the FQHE corresponds to introducing the
phase shift function $\lambda_\alpha \pi {\rm sgn} (k)$ in the
present 1D system, which can be given
by the $1/r^2$ interaction \cite{kawac}. Hence, one can clearly
see from  this correspondence why the family of $1/r^2$ models
can describe characteristic
properties of the hierarchical FQHE  remarkably.

\vskip 7mm
\noindent
{\it 5.6.  Dual bases: holon-spinon  and electron}
\vskip 2mm

We would like to point out another instructive
relationship to the FQHE \cite{kawad}. So far, we have
classified the  excitation
spectrum in terms of the matrix ${\bf T}$
and the corresponding quantum numbers out
of charge and spin  excitations. This
basis, which is referred to as {\it holon-spinon} basis,
always shows up when we use the Bethe ansatz method.
We should recall that there is an alternative
basis, i.e. {\it electron} basis.
Although the electron basis may not classify the excitation spectrum
when various excitations have different velocities,
it still describes the critical behavior of correlation functions
correctly \cite{kawad}. Quantum numbers in the electron basis
are obtained from those of holon-spinon basis via
a linear  transformation,
\begin{equation}
\vec N= {\bf U} \vec m, \hskip 5mm \vec J= ({\bf U}^t)^{-1} \vec d,
\end{equation}
with the matrix
\begin{equation}
{\bf U}=
\left(
\matrix {1  & -1      & \null  & \null   \cr
         \null         & 1       &  \ddots    & \null   \cr
        \null       & \ddots  & \ddots & -1  \cr
        \null       & \null   &  \null    & 1       \cr}
\right).
\end{equation}
The  matrix ${\bf T}$ is then  transformed
into the $\nu \times \nu$ symmetric matrix \cite{kawad},
\begin{equation}
\tilde {\bf T}= {\bf I} +{\bf V},
\end{equation}
where ${\bf I}$ is the $\nu \times \nu$  unit matrix.
Remarkably, we encounter the matrix ${\bf V}$ out
of the interaction parameters (see (77)). Hence the matrix ${\bf V}$
itself can classify the excitation spectrum
in the electron basis \cite{kawad}.   We should emphasize here that
there exists the exactly same
matrix  (97) in the 2D FQHE which
also characterizes the internal structure of
the hierarchical state  \cite{blok,zee}.
The above two kinds of bases are called the {\it symmetric}
basis and the {\it hierarchical} basis in the FQHE,
both of which characterize  the same topological order
of the hierarchical FQHE \cite{blok,zee}.

\vskip 5mm
\noindent
{\it 5.7.  Example: SU($\nu$) Sutherland model}
\vskip 2mm

As a simple example, we consider the first family of the
hierarchical models (76), i.e. the SU($\nu$)
Sutherland model characterized by
the parameters $V_{\alpha\beta}=\lambda$ \cite{ha,polya,hikami}.
In this case the exact wavefunction
has been obtained in the Jastrow form \cite{ha},
\begin{equation}
\psi= \prod_{l > m}|z_l^{(\alpha)}
-z_m^{(\beta)}|^{\lambda}
\psi_0  \ .
\end{equation}
Here $\psi_0$ is the wavefunction for
SU($\nu$) free electrons,
\begin{equation}
\psi_0 = \prod_{\alpha,j}[z_j^{(\alpha)}]^{2d_\alpha}
\prod_{\alpha,\beta,l>m}
(z_l^{(\alpha)}-z_m^{(\beta)})^{\delta_{\alpha_l \alpha_m}}
\exp[{i \over 2} \pi \, {\rm sgn} (\alpha_l -\alpha_m)].
\end{equation}
where $z_m^{(\alpha)}=\exp(2\pi i x_m^{(\alpha)}/L)$,
$L$ is the length of the periodic system and
$x_m^{(\alpha)}$ are spatial coordinates of
electrons with  spin $\alpha$ ($=1,2, \cdots, \nu$).
The current $2d_\alpha$ carried by $\alpha$-spin
electrons is assumed to take an integer value \cite{ha}.
This form of the  Jastrow wavefunction
clearly demonstrates the analogy to Jain's construction
of the wavefunction for the FQHE with  filling factor
$f_\nu=\nu/(\nu m+1)$ \cite{jain}. Namely, starting from
the noninteracting wavefunction, interacting electrons
can be described by introducing the Jastrow factor.
In the electron basis, the excitation is classified by the
symmetric matrix \cite{ha,kawad},
\begin{equation}
{\bf \tilde T}=
\left(
\matrix {1+\lambda  & \cdots      & \null  & 1+\lambda  \cr
         \vdots         & \ddots       &      & \vdots   \cr
        \null       &   & \ddots &   \cr
        1+\lambda   & \cdots   &      & 1+\lambda       \cr}
\right).
\end{equation}
In order to observe the symmetry property, it is more convenient
 to make use of the
 holon-spinon representation, which is characterized by
the following matrix \cite{kawad},
\begin{equation}
{\bf T}=
\left(
\matrix {1+\lambda  & -1      & \null  & \null   \cr
         -1         & 2       &  -1    & \null   \cr
        \null       & \ddots  & \ddots & \ddots  \cr
        \null       & \null   &  -1    & 2       \cr}
\right).
\end{equation}
{}From this expression we can see  U(1) symmetry for the charge sector
($T_{11}$) and  SU($\nu$) symmetry for the spin sector
($T_{\alpha\beta}$ for $\alpha \geq 2$). Particularly
the $(\nu-1) \times (\nu-1)$
matrix out of $T_{\alpha\beta}$ for $\alpha, \beta
\geq 2$ is the SU($\nu$) Cartan matrix
which characterizes the Lie algebra with SU($\nu$) symmetry.
  Hence,  the
critical behavior of the spin sector is
characterized by level-1 SU($\nu$) Kac-Moody algebra
with the central charge  $c=\nu-1$, and
the holon sector is governed by $c=1$ CFT
for which the scaling dimensions vary continuously
according to $\lambda$.

It is now easy to obtain
critical  exponents for the  correlation functions.
For example, the critical exponent for the
momentum distribution function is obtained as \cite{kawad}
\begin{equation}
\theta_\alpha ={1 \over 2} \nu \lambda^2(1+\nu \lambda)^{-1},
\end{equation}
which is in contrast to the chiral case,
\begin{equation}
\theta_\alpha^{(c)}= \lambda.
\end{equation}
Note that the critical exponent for the
chiral case exactly
coincides with the result for
chiral Luttinger liquids of the
edge states with
filling fraction $f_\nu=\nu/(\nu\lambda+1))$ \cite{wen}.

\vskip 15mm
\begin{center}
\begin{bf}
\S  6.   Confined Models and  Renormalized Harmonic Oscillators
\end{bf}
\end{center}
\vskip 3mm

We have been concerned so far with quantum models with
periodic boundary conditions. There is another class
of the integrable $1/r^2$ models
with {\it harmonic confinement}
 \cite{calo,sutha,brink,mp,frahm,kawag,kawah,vacek,vaceka}.
Besides much interest in the integrability
of the confined models \cite{mp},
there have also been several attempts to apply them to
conductance oscillations in  mesoscopic
systems \cite{tewari,jp,kawah,vacek}.
In this section, we  propose
a systematic construction of the energy spectrum
for the class of $1/r^2$ models
with harmonic confinement \cite{kawag}.
This approach is  referred to as the
{\it renormalized-harmonic oscillator} (RHO)
solution, since the essence of the idea is
that all the interaction effects are
incorporated in terms of the renormalized quantum numbers
of oscillators.  In this sense, this method
is regarded as  a variant of the ABA.
We use this idea for a systematic  construction of the spectrum
for the confined $1/r^2$ models \cite{kawag}.
By constructing the eigenfunctions explicitly \cite{vacek,vaceka},
we then give a proof that the RHO solution indeed provides
the exact spectrum of the
models. In the final part of this section we
briefly mention an application of the model to
conductance-oscillation phenomena in narrow channels
\cite{tewari,jp,vacek}.

\vskip  7mm
\noindent
{\it  6.1.  Renormalized-harmonic-oscillator solution}

\vskip 2mm
\noindent
{\it  6.1.1.
 Calogero-Sutherland model}

To depict the essence of the idea, we begin with
the Calogero-Sutherland model  which is given by interacting
spinless fermions (or bosons) confined by
harmonic potential \cite{calo,sutha},
\begin{equation}
H = - \frac{1}{2} \sum_{i=1}^{N} \frac{\partial^2}{\partial x_i^2}
    + \frac{1}{2} \sum_{i=1}^{N} \omega^2 x_i^2
    + \sum_{j>i} \frac{\lambda(\lambda+1)}{(x_j-x_i)^2},
\end{equation}
where the interaction parameter is assumed to be $\lambda \geq 0$.
Here, we recall the key property common to
$1/r^2$ models, i.e.  the  interaction gives rise to
the  repulsion among energy levels, and enlarges the spacing of
quantum numbers {\it uniformly}.  As seen in the
previous sections, the repulsion effect  can be
formulated by introducing the  step-wise
phase shift of the two-body $S$-matrix in the periodic case.
This is the heart of the ABA in which the renormalized
quantities correspond to
the rapidities in the ABA equations.

For the case of harmonic confinement,
the repulsion of energy levels can be taken into account
by the renormalization of the quantum
numbers for oscillators \cite{kawag}. In the  RHO
approach,  all the interaction effects are conjectured to be
incorporated into the {\it renormalized  quantum number},
$n_j$, and the energy is given in the expression for free harmonic
oscillators,
\begin{equation}
E= \omega \sum_{j=1}^{N} (n_j +{1\over 2}).
\end{equation}
We note that the renormalized quantum numbers
should be related to the conserved charges which ensure the
integrability of the model, although the explicit relation has
not been derived yet. As is the case for the periodic case,
the repulsion of  energy levels
are described by introducing the step function.
 Hence the quantum number $n_j$ is
to be determined by the equation \cite{kawag},
\begin{equation}
n_j=  I_j + \lambda \sum_{l =1}^N \theta (n_j- n_l),
\end{equation}
where $I_j (= 0, 1, 2, \cdots$) is the bare
quantum number, and the step function is introduced such that
$\theta(x)=1$ for $x>0$ and   $\theta(x)=0$ for $x \leq 0$.
Consequently we obtain  the energy for the
Calogero-Sutherland model as,
\begin{equation}
E=   \omega [ {1 \over  2} \lambda N(N-1) +
\sum_{j=1}^{N} (I_j+{1 \over 2})].
\end{equation}
The ground state is given by the successive quantum numbers
$I_j= 0, 1, \cdots, N-1$,
and  particle-hole excitations are described by changing
the quantum numbers $I_j$ from those for the ground state.
Remarkably enough, the above results deduced from
the RHO reproduce the exact spectrum
for the Calogero-Sutherland model
\cite{calo,sutha,brink,mp}.

\vskip  7mm
\noindent
{\it  6.1.2.  SU($\nu$) confined model }

The RHO solution can be applied to more general
multicomponent models \cite{kawag}.  As an example,
let us study the SU($\nu$) electron model with
harmonic confinement, which is a variant of the
SU($\nu$) Sutherland model with periodic boundary
conditions discussed in the previous sections.
The Hamiltonian reads
\cite{mp,frahm,vacek},
\begin{equation}
H = - \frac{1}{2} \sum_{i=1}^{N} \frac{\partial^2}{\partial x_i^2}
    + \frac{1}{2} \sum_{i=1}^{N} \omega^2 x_i^2
    + \sum_{j>i} \frac{\lambda(\lambda+P_{ij}^{\alpha\beta})}
{(x_j-x_i)^2},
\end{equation}
with   the spin-exchange operator $P_{ij}^{\alpha\beta}$
of two particles ($\alpha,\beta =1,2, \cdots, \nu$).
As usual, to complete the diagonalization for the SU($\nu$) model,
it  is necessary to introduce the
set  of renormalized quantum numbers $n_j^{(\alpha)}$
($\alpha=1,2, \cdots, \nu$) which
satisfy the nested algebraic equations \cite{kawag},
\begin{equation}
 n_j^{(1)} =   I_j^{(1)}
- \sum_{m}^{M_2} \theta(n_j^{(1)} -n_m^{(2)})
+ \lambda \sum_{l}^{M_1}\theta(n_j^{(1)}-n_l^{(1)}),
\end{equation}
\begin{equation}
\sum_{l}^{M_\alpha}\theta(n_m^{(\alpha)}-n_l^{(\alpha)}) +
I_{m}^{(\alpha)}
=  \sum_{j}^{M_{\alpha-1}} \theta(n_m^{(\alpha)}-n_j^{(\alpha-1)})
+\sum_s^{M_\alpha+1} \theta(n_m^{(\alpha)}-n_s^{(\alpha+1)}),
\end{equation}
for $2 \leq \alpha \leq \nu$. Here
the bare quantum numbers $I_j^{(\alpha)}$
are the non-negative integers
($=0, 1, \cdots$) which specify $\nu$ kinds of elementary excitations.
In the above equations, the quantity
$M_\alpha=\sum_{\beta=\alpha}^\nu N_\beta$ was introduced,
where $N_\beta$ is the number of electrons with $\beta$ spin
($M_1=N=\sum_{\beta=1}^{\nu} N_{\beta}$).

The  energy is written in the expression for  harmonic oscillators,
$E= \omega \sum_{j=1}^{N} (n_j^{(1)} +1/ 2)$, in the RHO method.
By substituting the nested equations to this formula, and
then iterating the substitutions,
we obtain the final expression for the energy as \cite{kawag},
\begin{equation}
E =  \omega [{1 \over 2} \lambda N(N-1)
 + \sum_{\alpha=1}^{\nu}
({1 \over 2} N_\alpha^2 - {1 \over 2} M_\alpha(M_\alpha-1)
+ \sum_{j=1}^{M_\alpha} I_j^{(\alpha)})].
\end{equation}
The ground state is described  by  the successive
non-negative quantum
numbers $I_j^{(\alpha)}= 0, 1, \cdots, M_\alpha-1$,
resulting in the ground-state energy
for SU($\nu$) singlet,
\begin{equation}
E_g=  \omega [
{1 \over 2} \lambda N(N-1)
 + {1 \over 2} \sum_{\alpha=1}^{\nu} N_\alpha^2].
\end{equation}
We note that all the interaction effects are
incorporated via the first term of (112),
and any effects of the interaction do not show up so far as
the number of electrons are fixed.
Hence, the excitation spectrum is described by
free oscillators in case of the fixed number of electrons.
We should note that this does not mean
the system to be out of free oscillators, as
is clearly seen from the
level-repulsion effects in the RHO equations.

\vskip  7mm\noindent
{\it 6.1.3.  hierarchical models}

It is straightforward to apply the RHO solution to
the hierarchical family of the confined models analogous to
(76) \cite{kawakura},
\begin{equation}
H = - \frac{1}{2} \sum_{i=1}^{N} \frac{\partial^2}{\partial x_i^2}
 + \frac{1}{2} \sum_{i=1}^{N} \omega_0^2 x_i^2
 + \sum_{j>i} \frac{V_{\alpha\beta}(V_{\alpha\beta}
+P_{ij}^{\alpha\beta})}{(x_j-x_i)^2},
\end{equation}
where the interaction parameters are
the same as those of (77).
Applying the RHO techniques describe above, we
obtain the energy as,
\begin{equation}
E/ =\omega  \sum_{\alpha=1}^{\nu}[
{1 \over 2} N_\alpha^2
+ {1 \over 2} (\lambda_\alpha-1)
 M_\alpha(M_\alpha-1)
+ \sum_{j=1}^{M_\alpha} I_j^{(\alpha)}].
\end{equation}
The ground-state energy then takes the form,
\begin{equation}
E_g = {\omega  \over 2} \sum_{\alpha=1}^\nu
[N_\alpha^2 +\lambda_\alpha M_\alpha(M_\alpha-1)].
\end{equation}
We can classify the excitation spectrum in the matrix
formula, which leads to the expression similar to
(92) in which we should replace $(2\pi v/L)$ by $\omega$.
The detail of this part will be reported elsewhere \cite{kawakura}.

\vskip  7mm
\noindent
{\it  6.2. Construction of eigenfunctions}

We have studied so far the energy spectrum
making use of  the RHO solution which is deduced
from the level-repulsion effects of $1/r^2$ interaction.
Here we give the microscopic foundation
of the RHO solution by constructing the eigenfunctions
explicitly.  We give a brief review of the
results of Vacek et al.  to construct the
eigenfunctions of the ground-state \cite{vacek}
as well as the excited states \cite{vaceka}.

\vskip 3mm
\noindent
{\it 6.2.1. ground state }

We start by writing down a general form of
Jastrow wavefunction which is to be the exact
eigenstate of the family of  $1/r^2$ models.
It consists of two parts,
\begin{equation}
{\mit \Psi}(x_1 \alpha_1, \ldots ,x_N \alpha_N) =
\prod_{j>i} |x_j-x_i|^\lambda
{{\mit \Psi}_0} (x_1 \alpha_1, \ldots ,x_N \alpha_N),
\end{equation}
where the first one is the Jastrow factor, and
the second, ${{\mit \Psi}_0}, $ is the eigenfunction
for the noninteracting SU($\nu$) electron model
($\alpha=1,2, \cdots, \nu$).
For the above  wavefunction,
therefore,  all the interaction effects are assumed to be
taken into account solely by the Jastrow factor $|x_j-x_i|^\lambda $.
We should like to emphasize that this expression is
quite general for the $1/r^2$ models.
For example, the wavefunction for the SU($\nu$)
Sutherland model takes this form (99). We will see
that this is also the case for excited states.

For the confined model of (108), the noninteracting
wavefunctions is given by that for
free electrons in the harmonic potential.
Hence, we can expect that  the ground-state
wavefunction  should take the form \cite{mp,vacek},
\begin{equation}
{\mit \Psi}_g =  \prod_{j>i} |x_j-x_i|^\lambda
(x_j-x_i)^{\delta_{\alpha_j \alpha_i}}
{\rm exp} \left [
i\frac{\pi}{2}{\rm sgn}(\alpha_j-\alpha_i)
\right ]
\prod_{i=1}^{N} {\rm exp}(-\frac{\omega}{2}x_i^2).
\end{equation}
It has been  shown that this wavefunction is indeed
the exact eigenfunction of the
confined model (108) \cite{vacek}.
Applying the kinetic term and the potential
term on the wavefunction, we obtain
\begin{equation}
\frac{1}{{\mit \Psi}_g}
\left [ \frac{1}{2} \sum_{i=1}^{N}
\left (-\frac{\partial^2}{\partial x_i^2}
+ \omega^2 x_i^2 \right )\right ] {\mit \Psi_g}
     = \frac{ \omega}{2 } \left [\lambda N ( N - 1 )
     + \sum_\alpha N_\alpha^2 \right ]- u ,
\end{equation}
where the term $u$ is given by
\begin{equation}
u = \sum_{k<\ell} \frac{\lambda(\lambda-1)}{(x_k-x_\ell)^2} +
\sum_{k<\ell} \frac{2 \lambda \delta_{\alpha_k \alpha_\ell}}
{(x_k-x_\ell)^2} +
\sum_{i \neq k \neq \ell}
\frac{\lambda \delta_{\alpha_i \alpha_k}}{(x_i-x_k)(x_i-x_\ell)}.
\end{equation}
On the other hand, the action of the interaction term
yields \cite{vacek}
\begin{eqnarray}
 \frac{1}{{\mit \Psi_g}}
\left[ \sum_{j>i} \frac{\lambda
(\lambda+P_{ij}^{\alpha\beta})}{(x_j-x_i)^2}
\right ] {\mit \Psi_g}
& =&
\sum_{k<\ell} \frac{\lambda(\lambda-1)}{(x_k-x_\ell)^2} +
\sum_{k<\ell} \frac{2 \lambda \delta_{\alpha_k \alpha_\ell}}
{(x_k-x_\ell)^2}  \nonumber \\
&+&
\sum_{k<\ell} \frac{\lambda}{(x_k-x_\ell)^2}
\left [1-\prod_{i \neq k\ell}
\left (\frac{x_i-x_\ell}{x_i-x_k} \right )
^{\delta_{{\alpha_i}{\alpha_k}}-\delta_{{\alpha_i}{\alpha_\ell}}}
\right ]
\left (1 - \delta_{\alpha_k \alpha_\ell} \right ).
\end{eqnarray}
A remarkable point is that multiparticle terms in the
above expressions (119) and (120)
coming from the kinetic energy and the interaction energy
cancel each other completely \cite{vacek}.
Consequently, it has been proven that
the Jastrow wavefunction (117)
is the exact eigenfunction of the Hamiltonian (108),
and the corresponding energy is given by the expression (115).
The detail of the calculation is given in ref. \cite{vacek}.
One can see evidence that the above eigenstate
indeed corresponds to the {\it ground state}, though it is not
easy to give a rigorous
proof for it. For instance, in the case of spinless fermions,
the wavefunction (117) reduces to the exact
ground-state wavefunction \cite{calo,sutha,polya}.
Also, in the limit of $\omega \rightarrow 0$,
we can show  that (117) is the ground-state wavefunction.
Based on these observations we believe
that the eigenfunction generally  describes
the exact {\it ground state} of the Hamiltonian.

\vskip 3mm
\noindent
{\it 6.2.2.  excited states}

Concerning the eigenfunctions for  excited states,
we can also use a general form of the Jastrow wavefunction
(116).  In this case, it is necessary to introduce the
exited wavefunctions of free electrons
for  ${{\mit \Psi}_0} $.
Therefore the following
Jastrow wavefunction should be
a candidate for the exact excited state \cite{vaceka},
\begin{equation}
{\mit \Psi} =    \prod_{j>i}
|x_j-x_i|^{\lambda}
\left [ \; \;\sum_{m_1+\ldots+m_N=I}\; \; \;
\prod_{i=1}^N \frac{1}{m_i !}
H_{m_i}(\sqrt{\frac{m \omega}{\hbar}} x_i)
\; \; \right ] \, {\mit \Psi}_g ,
\end{equation}
where $H_m$ is the Hermite polynomials
which can produce the excited states systematically.
It has been shown explicitly in \cite{vaceka} that this wavefunction
gives the exact excited state of the Hamiltonian (108) with
the corresponding energy obtained by the RHO method (114).
Therefore, the microscopic derivation of the eigenfunctions
establishes that the RHO approach proposed here gives
the exact solution to the confined $1/r^2$ systems.
We conclude this subsection by stressing that the general
form of the wavefunction (116) may be quite helpful for
constructing the excited states generally for the family
of the $1/r^2$ systems.

\vskip  10mm
\noindent
{\it 6.3. Application to conductance oscillations}

Here, we briefly mention some attempts to apply
the confined $1/r^2$ model to the transport of electrons
through a narrow channel of the semiconductor nanostructure.
We first note that it is crucial to take
into account the effects of  mutual electron
interactions to explain experiments of conductance oscillations
\cite{cond,coulomb}.
Also, it is known experimentally that one-dimensional electrons
are confined to a finite segment
by impurities or constrictions \cite{coulomb}.
Hence it may be necessary to introduce a  {\it interacting}
electron model with certain {\it confining potential}.
Motivated by the above experiments,
the $1/r^2$ models with harmonic confinement have been
applied to conductance oscillations \cite{tewari,jp},
and several characteristic properties have been explained.
Subsequently, the effect of the internal spin degrees
of freedom was taken into account correctly \cite{kawah,vacek},
and it was  demonstrated that there can be two kinds of periods
in the conductance oscillations.  We briefly summarize the
results  of refs. \cite{kawah,vacek}. We will be concerned with
the SU(2) case of (108) which corresponds to the ordinary
electron systems.

In order to consider the transport in narrow channels,
a weak coupling is introduced  between the segment
given by the model Hamiltonian (108) and
two reservoirs. Let us suppose that the conductance
is controlled by a resonant tunneling between the
one-dimensional segment and
the reservoirs.  Therefore  a peak in conductance
oscillations occurs when the chemical potential of the
reservoirs satisfy the relation,
$\mu(N)=E_g(N+1)-E_g(N)$. The spacing $\delta$
of two successive peaks in the conductance oscillations
is then given by $\delta(N)=\mu(N+1)-\mu(N)$.
Using the formula (115) for the ground state energy for the
SU(2) case (electrons), one can see that there appear
{\it two independent periods} of the conductance
oscillations \cite{kawah,vacek},
\begin{equation}
\delta_1 =  \omega \lambda, \hskip 5mm
\delta_2 =  \omega (\lambda+1),
\end{equation}
reflecting the exchange effect
due to  the internal spin degrees of freedom.
We stress that this result improves those
previously obtained by
Tewari \cite{tewari}  and also by Johnson and Payne \cite{jp},
who employed the {\it spinless} fermion models
and concluded a {\it single period} for
the conductance oscillations for any strength of the
interaction.

For the parameters  employed in ref. \cite{jp},
two periods become $\delta_1=7.5\omega$ and
$\delta_2=8.5\omega$ in the present model.
Hence, the correction due to the exchange effect is small
for these parameters (strong correlation regime).
The exchange effect, however, becomes more conspicuous
when the interaction becomes weaker (smaller $\lambda$),
and then  two periods becomes  more distinct
from each other.  In the weak-coupling limit
($\lambda \rightarrow 0$), the present model reproduces
the results for free electrons in the harmonic well:
$\delta_1 \rightarrow 0$, $\delta_2 \rightarrow  \omega$.
It may  be interesting to study experimentally
whether two periods due to the exchange effect
can be observed in the conductance oscillations.

\vskip 10mm
\begin{center}
\begin{bf}
\S 7. Summary
\end{bf}
\end{center}

We have reviewed our recent works
on the quantum  $1/r^2$ models.
We have successfully applied the ABA solution to
the multicomponent models
such as  the SU($\nu$) spin chain, the Osp($\nu,1$)
supersymmetric {\it t-J} model,  and
the hierarchical models related to the FQHE.  Applying
CFT techniques to the ABA equations,
we have studied the critical behavior
of this class of integral models.
As for the confined  models with harmonic potential,
the RHO solution has been proposed,
which enables us to construct the energy spectrum
systematically.
We have then proven, by explicitly constructing the
eigenfunctions, that the RHO indeed gives  the exact solution
to the family of the confined $1/r^2$ models.
The results have been used to discuss the exchange-correlation
effects on the conductance oscillations in
narrow channels.

\vskip 8mm
\begin{center}
\begin{bf}
Acknowledgements
\end{bf}
\end{center}
We would like to express our sincere thanks to
Y. Kuramoto, A. Okiji and  K. Vacek
and S.-K. Yang for useful discussions and fruitful
collaborations.  Valuable discussions with
I. Affleck, H. Frahm, D. Haldane, P. Horsch, A. Kl\"umper,
V. Korepin, A. Schadschneider,
S. Shastry and J. Zittartz are also acknowledged.
This work was partly supported by Grant-in-Aid from
the Ministry of Education, Science and Culture
and also by Monbusho International Scientific Research
Program.


\end{document}